\documentstyle[twocolumn,prl,aps]{revtex}  
\begin{document}
\draft
\title{Kinetics and Jamming Coverage in a\\
Random Sequential Adsorption of Polymer Chains}

\author{Jian-Sheng Wang and Ras B. Pandey\cite{Pandey-Permanent-Address}}

\address{Department of Computational Science,
National University of Singapore,
Singapore 119260, Republic of Singapore}

\date{6 May 1996}
\maketitle

\begin{abstract}

Using a highly efficient Monte Carlo algorithm, we are able to
study the growth of coverage in a random sequential adsorption
(RSA) of self-avoiding walk (SAW) chains for up to $\sim
10^{12}$ time steps on a square lattice.  For the first time,
the true jamming coverage ($\theta_J$) is found to decay with
the chain length ($N$) with a power-law $\theta_J \propto
N^{-0.1}$.  The growth of the coverage to its jamming limit can
be described by a power-law, $\theta(t) \approx \theta_J -c/t^y$
with an effective exponent $y$ which depends on the chain
length, i.e., $y \simeq 0.50$ for $N=4$ to $y \simeq 0.07$ for
$N=30$ with $y \to 0$ in the asymptotic limit $N \to \infty$.

\end{abstract}


Studying the kinetics of random sequential adsorption (RSA) has
attracted a considerable interest in recent years
\cite{Evans-93,%
Feder-80,%
Pomeau-80,%
Swendsen-81,%
Schaaf-group-89,%
Vigil-Ziff-89,%
Viot-Tarjus-90,%
Dickman-Wang-Jensen-91,%
Svrakic-Henkel-91,%
Brosilow-Ziff-Vigil-91,%
Tarjus-Viot-91,%
Tarjus-Talbot-91,%
Meakin-Jullien-92,%
Viot-Tarjus-Ricci-Talbot-92,%
Wang-Privman-Nielaba-93,%
Becklehimer-Pandey-94,%
Sinkovits-Pandey-94,%
Baram-Fixman-95} because of its enormous applications
\cite{Flory-39,%
Finegold-Donnell-79,%
Wolf-Burgess-Hoffman-80,%
Onoda-Liniger-86,%
Privman-Frisch-et-al-91,%
Ramsden-et-al-1993}
in the adsorption processes involving a variety of species from
a point-like particle to a protein-like complex structure in
physical, chemical, and biological systems.  Some of the examples
include binding of ligands on polymer chains, coating, designing
composites, chemisorption, physisorption, and reaction of
molecular species including globular protein on surfaces and
interfaces, etc.  These adsorption
processes may be divided into two categories: (1)
annealed adsorption where the species are mobile (a thermal
equilibration for the interacting adsorbants) before they settle
onto the surface -- a cooperative sequential adsorption.  (2)
Quenched adsorption where the adsorption occurs without
subsequent diffusion or desorption.  We consider the latter
category known as random sequential adsorption.

The problem of RSA in one dimension
\cite{Flory-39,%
Renyi-58,%
Gonzalez-Hemmer-Hoye-74,%
Bartelt-Privman-review-91}
is well understood with exact results for some adsorption
processes.  Understanding the growth of coverage in two
dimensions with the RSA lacks rigorous results by analytical
methods due to their intractabilities especially for objects
with polydisperse shapes.  Therefore, computer simulations
remain one of the primary tools to investigate these problems.
Numerical results \cite{Feder-80} and theoretical analyses
\cite{Pomeau-80,Swendsen-81} for the deposition of
disks on continuum suggest that the coverage follows the Feder's
law \cite{Feder-80} at large time,
\begin{equation}
\theta(t) \approx \theta_J - {c \over t^{1/d} }, \label{eq-Feder-law}
\end{equation}
where, $\theta(t)$ is the coverage at time $t$, $\theta_J$ is
the jamming coverage, $d$ is the dimensionality of the host
space, and $c$ is a constant.

The temporal dependence, Eq.~(\ref{eq-Feder-law}), seems
to describe the deposition of hyperspheres.  For squares
\cite{Viot-Tarjus-90}, rectangles \cite{Vigil-Ziff-89}, and
ellipses \cite{Viot-Tarjus-Ricci-Talbot-92}, on the other hand,
the above law is still valid but $d$ is replaced by $d_f$, the
number of degrees of freedom of the corresponding objects.  For
disks with polydisperse sizes the computer simulation studies
\cite{Meakin-Jullien-92} seem consistent with the theoretical
prediction \cite{Tarjus-Talbot-91}, $\theta(t) \approx \theta_J
- c\, t^{-y}$, where $y = 1/(d+1)$.  As a natural extension, one
would like to consider objects not only with polydisperse 
sizes but also with
polydisperse shapes, a more complex problem in the studies
of RSA.  Therefore, we perform a large-scale Monte Carlo
simulation to study the adsorption of polymer chains which have
polydispersity in both shapes and sizes.  A model chain is one
of the most simple ramified objects with its well-known shape
and size distributions in a variety of systems
\cite{Flory-69,de-Gennes-79}.  The RSA of polymer chains is
nevertheless relevant in applications such as coating and paint.
In contrast to theoretical predictions for the RSA of
polydisperse objects of regular shapes, we find that an
effective exponent $y$ depends on the chain length.
Furthermore, we are able to reach the true jamming coverage
($\theta_J$) with an efficient algorithm which enables us to
predict a power-law dependence on the chain length ($N$),
$\theta_J \propto N^{-0.1}$.

We consider an $L \times L$ square lattice with a periodic
boundary condition.  A polymer chain is modeled by a
self-avoiding random walk (SAW) which is generated on the trail
of a non-reversal random walk (NRRW) with self-avoiding
constraints.  The chains are dropped onto the square lattice,
one at a time sequentially.  If a chain overlaps with previously
deposited chains, the attempt is rejected.  Once the chain is
deposited on the lattice, it sticks on the surface permanently.
The deposition rate or equivalently the time scale is fixed with
the following algorithm.  In unit time $\Delta t = 1$, $L^2$
attempts are made to generate and deposit SAW chains each of
length $N$ starting from a randomly selected site.  As soon as a
walk overlaps with the previously deposited site (from other
chains or itself), the walk is abandoned, and a new attempt is
made to deposit a chain starting from a randomly chosen new
site.

Due to the large number of conformational states of polymer
chains, the deposition becomes very slow in the late stage.  
An event-driven method
\cite{Brosilow-Ziff-Vigil-91,Wang-94,Bortz-Kalos-Lebowitz-75}
is, therefore, used to speed up the simulation.  This is
accomplished by identifying the early part of growing chains,
i.e.,  the partial chains and then classifying them periodically as
``available'' or ``not available'' for the deposition in an
iterative fashion.  The subsequent depositions are then made 
starting from the available partial chains choosing at random.
The list of the partial chains is limited by the
available computer memory and governs the speed of the program.
It appears that there is an optimal list size for a given value
of $N$.

Now, in order to maintain the same dynamics, each deposition
attempt does not simply advance the time by $\delta t = L^{-2}$
as before, but by an amount $j \delta t$, where $j$ is a random
integer with an exponential distribution, $p(1-p)^{j-1}$.  $p$
is the ratio of the number of potentially available partial
chains in the list to the total number of chains of partial
length $n$, $L^2 Z_n^{nrrw}$, where $Z_n^{nrrw}$ is the number
of NRRW chains of length $n$.  The random integer can be
generated by
\begin{equation}
j = 1 + \left\lfloor { \ln \xi \over \ln (1-p) } \right\rfloor,
\end{equation}
where $\xi$ is a uniformly distributed random number between 0
and 1.

The simulation is performed on IBM SP2 and fast workstation
clusters.  The total amount of CPU time for the computation is
about six months equivalent of a single DEC AlphaStation 250/266.
Variation of the coverage $\theta$ with time $t$ is presented
in Fig.~\ref{fig-1} for various chain lengths.  We immediately
notice that a rapid increase in the coverage in the short time
regime is followed by a very slow growth in the long time.  We
can divide the characteristic behavior in three time regimes:
(1) short time regime, (2) intermediate regime, and (3) very
late-stage regime.

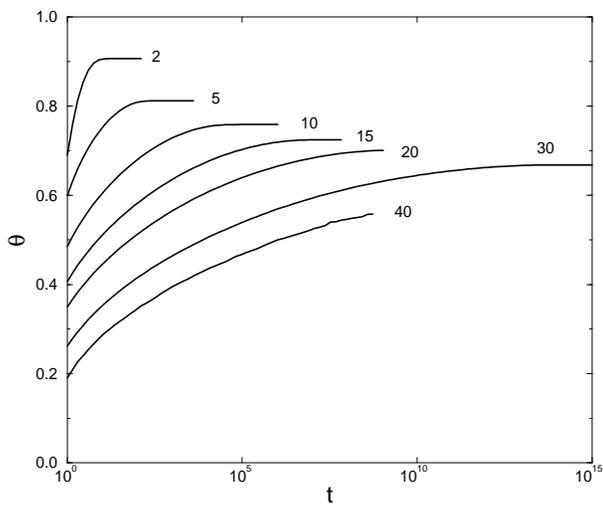
\begin{figure}
\caption{Semi-logarithmic plot for the difference between jamming
coverage and coverage at time $t$, for chain lengths $N=2$, 5, 10,
and 15.  For a better view of all the curves on the same
plot, we plotted again the normalized time $t/\tau$.  The straight
line is the pure exponential decay $e^{-t}$ for $N=1$.  The insert
shows $\tau$ as a function of $N$, the straight line is the 
theoretical prediction, Eq.~(\ref{eq-tau}).}
\label{fig-2}
\end{figure}

The short and very long time regimes (1) and (3) are easily
understood.  The coverage at very short time is proportional to
$t$, since the depositions of nearly all SAWs are accepted.  In
very late stage \cite{Bartelt-Privman-91}, the dynamics is
controlled by filling independently the last pores.  The holes
disappear according to $e^{-t/\tau}$, where $1/\tau$ is the
probability that a given hole is being filled (in unit time).
For the deposition of chains, the slowest mode is filling the
void in exactly two ways (out of $Z_N^{nrrw}$ possibilities).
Thus
\begin{equation}
\tau = {Z_N^{nrrw}\over 2} = 2 \cdot 3^{N-2}, \qquad N > 1.
\label{eq-tau}
\end{equation}
The exponential decays and the prediction for $\tau$ agree very
well with simulation data, see Fig.~\ref{fig-2}.

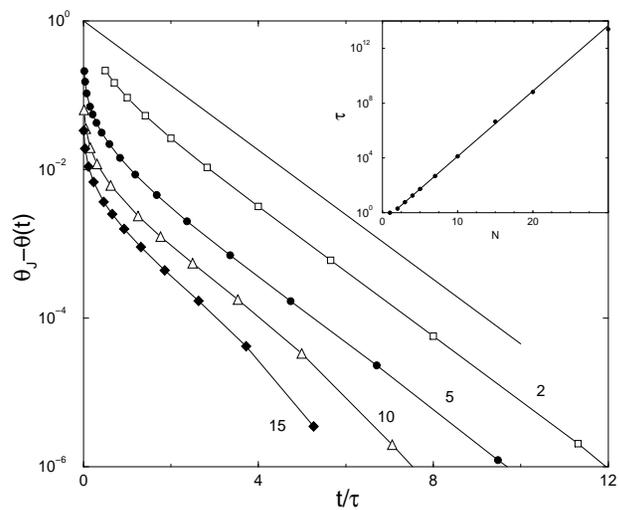
\begin{figure}
\caption{Derivative of coverage vs.~time on a log-log scale.
The number indicates chain length $N$.  Insert shows the
variation of exponent $y$ with time evaluated at a regular
interval for $N=30$.  The statistics is presented in
table~\ref{tb-1}.}
\label{fig-3}
\end{figure}

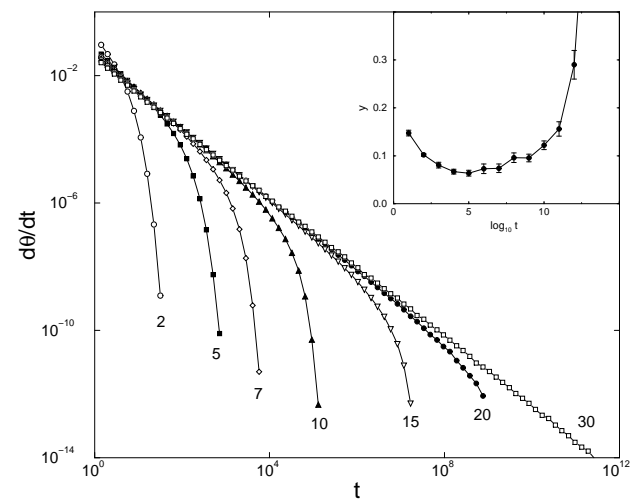
\begin{figure}
\caption{Exponent $y$ obtained by least-squares fits of
Fig.~\ref{fig-3}, plotted against inverse chain length.}
\label{fig-4}
\end{figure}

In the intermediate time regime (2), $ 1 < t \ll \tau$, a
power-law dependence is observed even on discrete lattice i.e.,
\begin{equation}
\theta(t)  \approx \theta_J - c\, t^{-y},
\end{equation}
with an effective exponent $y$.  Since the jamming limit
$\theta_J$ is not known accurately for long chains, it is
instructive to consider the derivative of $\theta$ with respect
to $t$,
\begin{equation}
{d \theta \over dt} \propto {1 \over t^{1+y}}. \label{eq-power-law}
\end{equation}
This quantity is plotted in Fig.~\ref{fig-3} for various chain
lengths.  We see a crossover from a power-law
variation of the rate of coverage in the intermediate time
regime to an exponential decay in the long time especially for
short chains.  Note that this
intermediate regime expands on increasing the chain length.  We
observe a remarkable power-law behavior over 12 decades in time
with the largest chain length $N=30$ for which the true jamming
limit is reached in our simulations.  Least-square fits in the
power-law regime give a reliable estimate of the effective
exponent $y$, particularly for $N \le 30$.  Note that the small
value of $y \simeq 0.07$ for $N=30$ 
is stable over a relatively large time
scale within the statistical fluctuations (see the inset in
Fig.~\ref{fig-3}, and Fig.~\ref{fig-4}). 
A logarithmic fit $\theta(t) \approx
\theta_J - c/\log(t)$ appears less satisfactory.  We see that the
magnitude of the exponent $y$ depends systematically on the
chain length, $y \simeq 0.50$ for $N=4$ to $y \simeq 0.07$ for
$N=30$.  To our knowledge, none of the previous studies has
shown a size-dependent exponent $y$.  
A crude extrapolation (see Fig.~\ref{fig-4}) leads to a small
value of $y$ for larger chain length with a possibility for $y
\to 0$ as $N \to \infty$ within the statistical error.  

Let us recall the well-known Swendsen's argument
\cite{Swendsen-81} for Feder's law [Eq.~(\ref{eq-Feder-law})]
for the adsorption of disks on continuum.  At late stage,
the pore vanishes according to $e^{-kt}$, where $k \propto l^d$,
$l$ is the linear size of the pore, and $d$ is the spatial 
dimension.  Assuming an uniform linear
size $l$, the probability distribution for
the rate $k$ is $p(k)dk \propto 1 \cdot dl \propto k^{1/d - 1}
dk$.  Total contribution to the approach to jamming is then
\begin{equation}
\theta_J - \theta(t) \propto \int_0^{\infty} k^{{1\over d} - 1} 
e^{-kt} dk \propto t^{-1/d}.
\end{equation}
Viot and Tarjus \cite{Viot-Tarjus-90} generalized the above
result to anisotropic objects and concluded that the approach to
jamming is $t^{-1/d_f}$ for monodisperse objects, where $d_f$ is
the degrees of freedom of the objects.  For the objects with
regular shape but polydisperse sizes
\cite{Tarjus-Talbot-91}, $ \theta_J - \theta(t)
\propto t^{-1/(d_f+1)}$.

Polymer chains have much more internal degrees of freedom.  For
sufficiently long chains, the lattice structure becomes less
important.  The chains can be specified by the orientation of
their segments on a coarse-grain level.  The number of degrees
of freedom is proportional to the number of such segments.  Thus,
we may identify $d_f \propto N$ and consequently $y \propto
1/N$.  In fact the data for $y$ can be roughly characterized by
$y \approx 2/N$.

Alternatively, one may identify the degree of freedom as the number of
SAW configurations $Z_N^{saw}$ of length $N$, i.e., $d_f
\propto Z_N^{saw} \propto N^{\gamma - 1} z_{eff}^N$ where
$\gamma$ is a critical exponent and $z_{eff}
\le z-1$, where $z$ is the coordination number
\cite{Binder-Heermann}.
Then, the exponent $y \approx 1/(Z_N^{saw}+1)$ becomes much smaller
than our estimates. Thus, the application of Tarjus and Talbot result
\cite{Tarjus-Talbot-91} is not
valid if $d_f = Z_N^{saw}$ is assumed.  However, in asymptotic
limit for large $N$, $y \to 0$ as $N \to \infty$ is
consistent with the prediction of Tarjus and Talbot.

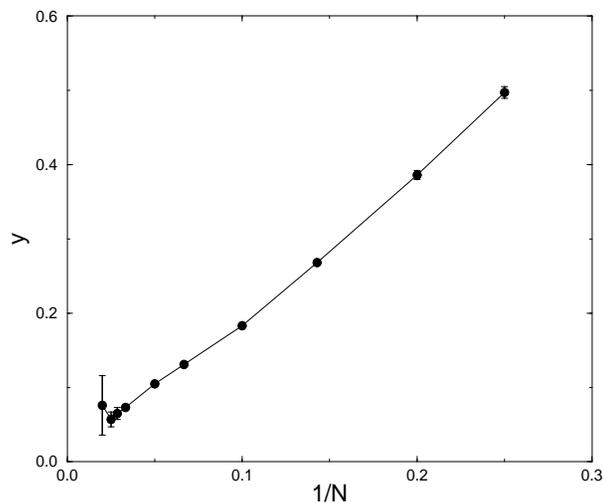
\begin{figure}
\caption{Jamming coverage versus chain length on a logarithmic scale.}
\label{fig-5}
\end{figure}

With the event-driven method we are able to reach jamming much
faster which gives us accurate estimates for the jamming
coverage $\theta_J$, see Table~\ref{tb-1}.  The jamming coverage
decreases with the chain length.  It is not clear {\sl a priori}
that the jamming coverage $\theta_J(N)$ goes to zero as $N$
approaches infinity.  However, assuming,
\begin{equation}
\theta_J(N)  \propto N^{-x}, 
\end{equation}
a $\theta_J(N)$ versus $N$ plot on a log-log scale (see 
Fig.~\ref{fig-5}) leads to a very small exponent $x \approx 0.11$.  
Note that the magnitude of the exponent $x$ is much smaller than
that of ref.~\cite{Becklehimer-Pandey-94} ($x \approx 1/3$) which
deals with the percolation \cite{Stauffer-Aharony-94} of chains.
The jamming coverage in the percolation, i.e., the maximum value
of the volume fraction is the ``first jamming coverage'' which
is far from the asymptotic limit of our ``true jamming
coverage'' here.

We have also analyzed the conformation of the chains during the
deposition process.  For SAW, it is well-known
\cite{Flory-69,de-Gennes-79} that the radius of gyration or
end-to-end distance varies with chain length as
\begin{equation}
R_g \propto R_e \propto N^{3/4}. 
\end{equation}
The question now is whether this law is modified for the chains
deposited on the lattice.  Our numerical data indicate that the
average radius of gyration is less sensitive to packing.  One of
the possible reasons for not observing much change in the
power-law dependence from that of a free SAW, is the fact that the
conformations of the chains in their early stage of deposition
dominate the average radius of gyration.  However, due to the
irreversible nature of the problem, the chains have memory of
their arrival time in the sense that their conformation and
correlation depend on $t$.  In fact, we do observe that the
radius of gyration decreases systematically with time especially
in the long time regime.

We thank Gan Chee Kwan, and Grace M Foo for useful discussions.
This work is supported in part by the Faculty Research Grant
RP950601.  Part of the computation was performed on IBM SP2 of
the National Supercomputing Research Centre. Warm hospitality
during his sabbatical visit at the NUS is acknowledged by RBP.

\begin{figure}
\caption{Coverage as a function of time, for chain length
$N$ indicated by the number. See table~\ref{tb-1} for the statistics
on the sample size and the number of independent runs.}

\label{fig-1}
\end{figure}

\begin{table}
\caption{Jamming coverage $\theta_J = \theta(t\to\infty)$ for various 
chain lengths.  Statistical errors on the last digits are indicated by 
the numbers in the parentheses.}
\label{tb-1}
\bigskip
\begin{tabular}{rlrr}
$N$  &  $\theta_J$(error) &   $L$   & number of runs \\
\hline
1 & 1 & & \\
2 & 0.906820(2)  &   1000  &   26200 \\
3 & 0.858296(4)  &    500  &   34000 \\
4 & 0.837055(13) &    500  &   14000 \\
5 & 0.81235(1)   &    500  &    7501 \\
7 & 0.78558(2)   &    500  &    2320 \\
10& 0.75895(1)   &    500  &    4701 \\
15& 0.72473(8)   &    500  &     700 \\
20& 0.70178(10)  &    500  &      33 \\
30& 0.6683(4)    &    500  &      28 \\
40& 0.654(9)     &    200  &       7 \\
\end{tabular}
\end{table}


\begin{references}
\bibitem[*]{Pandey-Permanent-Address} Permanent address: The Program in 
Scientific Computing, Department of Physics and Astronomy,
University of Southern Mississippi, Hattiesburg, MS 39406, USA.

\bibitem{Evans-93} For a review, see: J. W. Evans, {Rev. Mod. Phys.}
{\bf 65}, 1281 (1993).

\bibitem{Feder-80} J. Feder, {J. Theor. Biol.} {\bf 87}, 237 (1980).

\bibitem{Pomeau-80} Y. Pomeau, {J. Phys. A} {\bf 13}, L193 (1980).

\bibitem{Swendsen-81} R. H. Swendsen, {Phys. Rev. A} {\bf 24}, 504 (1981).

\bibitem{Schaaf-group-89} P. Schaaf and J. Talbot, {\sl Phys. Rev. Lett.} 
{\bf 62}, 175 (1989);
J. Talbot, G. Tarjus, and P. Schaaf, {Phys. Rev. A} {\bf 40},
4808 (1989).

\bibitem{Vigil-Ziff-89} R. D. Vigil and R. M. Ziff, 
{J. Chem. Phys.} {\bf 91}, 2599 (1989).

\bibitem{Viot-Tarjus-90} P. Viot and G. Tarjus, {Europhys. Lett.} 
{\bf 13}, 295 (1990).

\bibitem{Dickman-Wang-Jensen-91} R. Dickman, J.-S. Wang, and I. Jensen,
{J. Chem. Phys.} {\bf 94}, 8252 (1991).

\bibitem{Svrakic-Henkel-91} N. M. \v Svraki\'c and M. Henkel, 
{J. Phys. (Paris)} {\bf I1}, 791 (1991).

\bibitem{Brosilow-Ziff-Vigil-91} B. J. Brosilow, R. M. Ziff, 
and R. D. Vigil, {Phys. Rev. A} {\bf 43}, 631 (1991).

\bibitem{Tarjus-Viot-91} G. Tarjus and P. Viot, {Phys. Rev. Lett.}
{\bf 67}, 1875 (1991).

\bibitem{Tarjus-Talbot-91} G. Tarjus and J. Talbot, 
{J. Phys. A} {\bf 24}, L913 (1991).

\bibitem{Meakin-Jullien-92} P. Meakin and R. Jullien, {Phys. Rev. A} 
{\bf 46}, 2029 (1992); {\sl Physica A} {\bf 187}, 475 (1992).

\bibitem{Viot-Tarjus-Ricci-Talbot-92} P. Viot, G. Tarjus, S. M. Ricci,
and J. Talbot, {J. Chem. Phys.} {\bf 97}, 5212 (1992); 
S. M. Ricci, J. Talbot, G. Tarjus, and P. Viot, {J. Chem. Phys.}
{\bf 97}, 5219 (1992).

\bibitem {Wang-Privman-Nielaba-93} V. Privman, J.-S. Wang, and 
P. Nielaba, {Phys. Rev. B} {\bf 43}, 3366 (1990);
J.-S. Wang, V. Privman, and P. Nielaba, {Mod. Phys. Lett.}
{\bf B7}, 189 (1993); J.-S. Wang, P. Nielaba, and V. Privman, 
{Physica A} {\bf 199}, 527 (1993).

\bibitem{Becklehimer-Pandey-94} J. L. Becklehimer and 
R. B. Pandey, {J. Stat. Phys.} {\bf 75}, 765 (1994).

\bibitem{Sinkovits-Pandey-94} R. S. Sinkovits and R. B. Pandey, 
{J. Stat. Phys.} {\bf 74}, 457 (1994).

\bibitem{Baram-Fixman-95} A. Baram and M. Fixman, {J. Chem. Phys.} 
{\bf 103}, 1929 (1995).

\bibitem{Flory-39} P. J. Flory, {J. Am. Chem. Soc.} {\bf 61}, 1518 (1939).

\bibitem{Finegold-Donnell-79} L. Finegold and J. T. Donnell, {Nature}
{\bf 278}, 443 (1979).

\bibitem{Wolf-Burgess-Hoffman-80} N. O. Wolf, D. R. Burgess, 
and D. K. Hoffman, {Surf. Sci.} {\bf 100}, 453 (1980). 

\bibitem{Onoda-Liniger-86} G. Y. Onoda and E. G. Liniger, 
{Phys. Rev. A} {\bf 33}, 715 (1986).

\bibitem{Privman-Frisch-et-al-91} V. Privman, H. L. Frisch, N. Ryde, 
and E. Matijevi\'c, {J. Chem. Soc. Faraday Trans.} {\bf 87}, 1371 (1991). 

\bibitem{Ramsden-et-al-1993} J. J. Ramsden, {Phys. Rev. Lett.} 
{\bf 71}, 295 (1993); J. J. Ramsden, G. I. Bachmanova, 
and A. I. Archakov, {Phys. Rev. E} {\bf 50}, 5072 (1994).

\bibitem{Renyi-58} A. R\'enyi, {Publ. Math. Inst. Hung. Acad. Sci.} 
{\bf 3}, 109 (1958).

\bibitem{Gonzalez-Hemmer-Hoye-74} J. J. Gonzalez, P. C. Hemmer, and 
J. S. H\o ye, {Chem. Phys.} {\bf 3}, 228 (1974).

\bibitem{Bartelt-Privman-review-91} M. C. Bartelt and V. Privman, 
{Int. J. Mod. Phys. B} {\bf 5}, 2882 (1991).

\bibitem{Flory-69} P. J. Flory, {Statistical Mechanics of Chain Molecules}
(Interscience, New York, 1969).

\bibitem{de-Gennes-79} P. G. de Gennes, {Scaling Concepts in 
Polymer Physics} (Cornell University Press, Ithaca, NY, 1979).


\bibitem{Wang-94} J.-S. Wang, {Inter. J. Mod. Phys. C} {\bf 5}, 707 (1994).

\bibitem{Bortz-Kalos-Lebowitz-75} A. B. Bortz, M. H. Kalos, 
and J. L. Lebowitz, {J. Comp. Phys.} {\bf 17}, 10 (1975).

\bibitem{Bartelt-Privman-91} M. C. Bartelt and V. Privman, 
{Phys. Rev.} {\bf A44}, R2227 (1991).

\bibitem{Binder-Heermann} K. Binder and D. W. Heermann,
{Monte Carlo Simulation in Statistical Physics}, p.~10,
(Springer-Verlag, Berlin, 1988).

\bibitem{Stauffer-Aharony-94} D. Stauffer and A. Aharony, 
{Introduction to Percolation},
2nd Ed., (Taylor \& Francis, London, 1994).

\end{references}
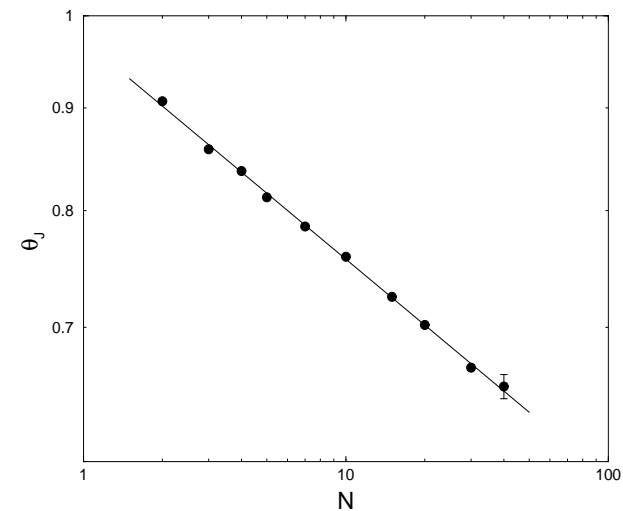
\end{document}